\documentclass[a4paper,11pt]{article}
\usepackage{pos}
\usepackage{aas_macros}

\graphicspath :
\graphicspath{{Figures/}}

\newcounter{affil}
\newenvironment{affil}[1]
{
 \noindent
  \refstepcounter{affil}%
    \textsuperscript{\theaffil}%
}%

\title{ The Giant Radio Array for Neutrino Detection (GRAND) Project}
 \ShortTitle{The GRAND Project}

\author*[a,b]{Kumiko Kotera}

\affiliation[a]{Sorbonne Universit\'{e} et CNRS, UMR 7095, Institut d'Astrophysique de Paris, 98 bis bd Arago, 75014 Paris, France}

\affiliation[b]{Vrije Universiteit Brussel (VUB), Dienst ELEM, Pleinlaan 2, B-1050, Brussels, Belgium
}

\emailAdd{kotera@iap.fr}
\forColl{GRAND}

\abstract{The GRAND project aims to detect ultra-high-energy neutrinos, cosmic rays and gamma rays, with an array of $200,000$ radio antennas over $200,000\,{\rm km}^2$, split into $\sim 20$ sub-arrays of $\sim 10,000\,{\rm km}^2$ deployed worldwide. The strategy of GRAND is to detect air showers above $10^{17}$\,eV that are induced by the interaction of ultra-high-energy particles in the atmosphere or in the Earth crust, through its associated coherent radio-emission in the $50-200$\,MHz range. In its final configuration, GRAND plans to reach a neutrino-sensitivity of $\sim 10^{-10}\,{\rm GeV}\,{\rm cm}^{-2}\,{\rm s}^{-1}\,{\rm sr}^{-1}$ above $5\times 10^{17}$\,eV combined with a sub-degree angular resolution. GRANDProto300, the 300-antenna pathfinder array, is planned to start data-taking in 2021. It aims at demonstrating autonomous radio detection of inclined air-showers, and study cosmic rays around the transition between Galactic and extra-Galactic sources. We present preliminary designs and simulation results, plans for the ongoing, staged approach to construction, and the rich research program made possible by the proposed sensitivity and angular resolution.}

\FullConference{37$^{\rm{th}}$ International Cosmic Ray Conference (ICRC 2021)\\
		July 12th -- 23rd, 2021\\
		Online -- Berlin, Germany}

\begin{document}
\maketitle

GRAND is a proposed large-scale observatory designed to discover and study the sources of ultra-high-energy cosmic rays (UHECRs). GRAND will detect the radio signals made in the Earth's atmosphere by ultra-high-energy (UHE) cosmic rays, gamma rays, and neutrinos. The sub-degree angular resolution of GRAND will make possible the discovery of the first point sources of UHE neutrinos. We present the detection concept, the expected performances, and the rich science case of the experiment, as well as the different stages planned to achieve the ultimate array.

\section{Detection concept}

GRAND is designed to detect {\it inclined} extensive air-showers (EAS) produced by UHE cosmic particles \cite{GRAND20}. While entering the atmosphere, UHE particles produce EAS, which in turn generate electromagnetic emission, mainly through the deflection of charged particles in the shower by the geomagnetic field. UHE tau neutrinos can also produce such electromagnetic signals by interacting with the Earth crust, giving birth to a tau lepton that can typically traverse a few tens of km of rock before exiting in the atmosphere and decaying, hence generating an Earth-skimming EAS. The geomagnetic emission is coherent in the 10s of MHz frequency range, generating short ($< 1\,\mu$s), transient radio pulses, with high enough amplitudes for the detection of showers with energy $\gtrsim 10^{16.5}\,$eV \cite{Huege:2016veh,SCHRODER20171}. 

GRAND will build on the mature radio-detection experience of past and existing radio-detection experiments (AERA, CODALEMA, LOFAR, TREND, Tunka-REX).
These instruments have focused on vertical EAS. Because of relativistic effects, the
radio emission is strongly beamed forward, with an opening angle corresponding to the Cherenkov
angle $\theta \lesssim 1^\circ$.
For vertical EAS, the radio-signal propagates over the $\sim 10\,$km thickness of the atmosphere, and leads to a footprint on the ground of few 100s\,m$^2$. The sampling of such a signal necessitates a dense radio array. For very inclined air-showers on the other hand, the radio emission can propagate over several tens of kilometers, inducing a footprint on the ground of several squared kilometers, which can be sampled with a sparse (kilometer-step) array. 

Radio antennas are ideal components to build giant arrays, being cheap, robust and scalable. In its final configuration, GRAND will consist of of $200,000$ antennas over $200,000\,{\rm km}^2$, split into $\sim 20$ sub-arrays of $10,000$ antennas located in different locations across the Earth. The locations of the sub-arrays will be chosen in radio-quiet environments with relatively easy access, and favorable topographies. An ideal topography consists of two opposing mountain ranges, separated by a few tens of kilometers. One range acts as a target for neutrino interactions, while the other acts as a screen on which the ensuing radio signal is projected. Simulations show that ground topographies inclined by few degrees only induce detection efficiencies typically three times larger than those obtained for flat areas \cite{Decoene_2021}.

\section{Expected Performances}\label{section:performances}

\begin{figure}[t]
\centering
\includegraphics[width=0.48\linewidth]{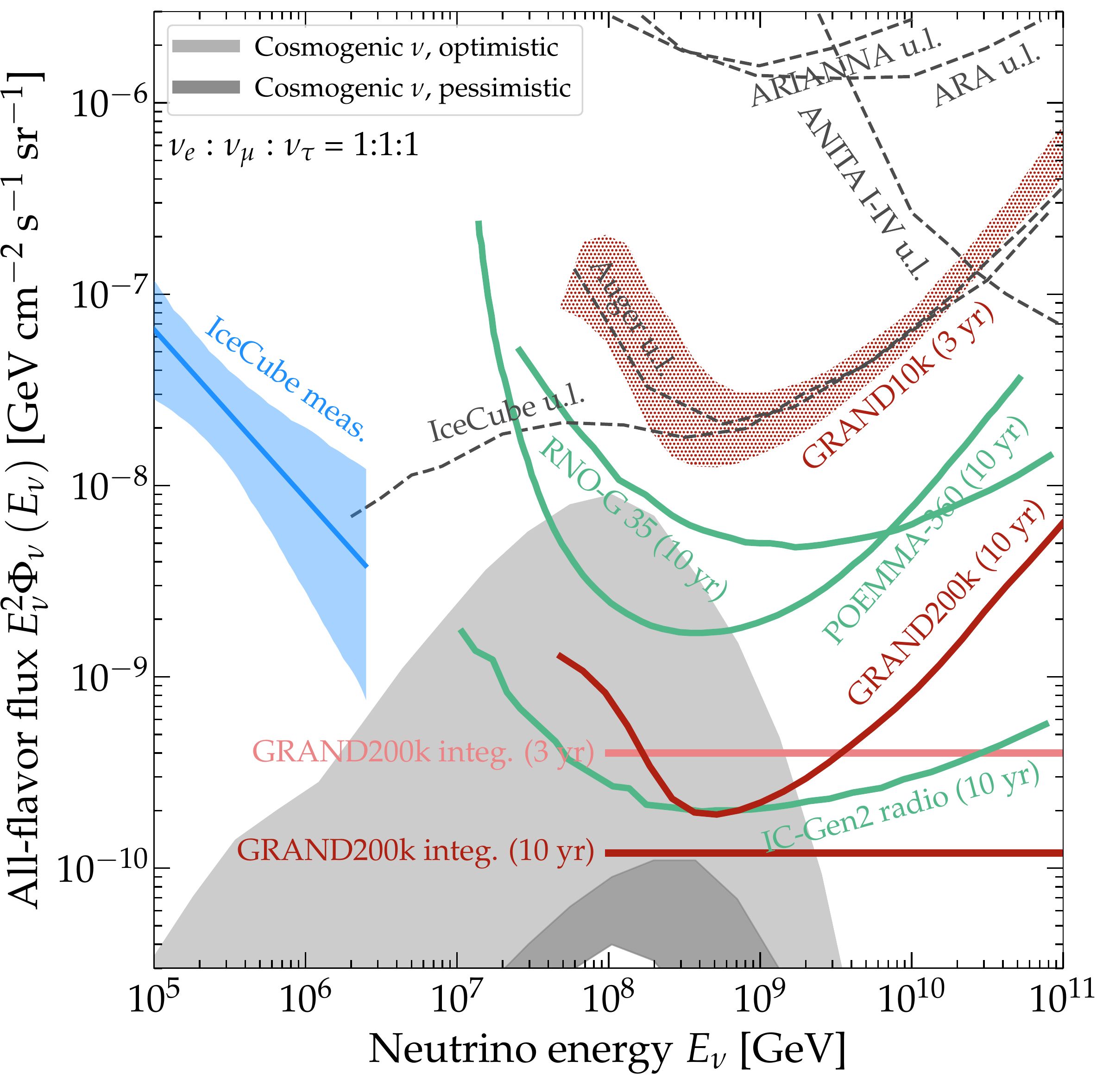}
\includegraphics[width=0.48\linewidth]{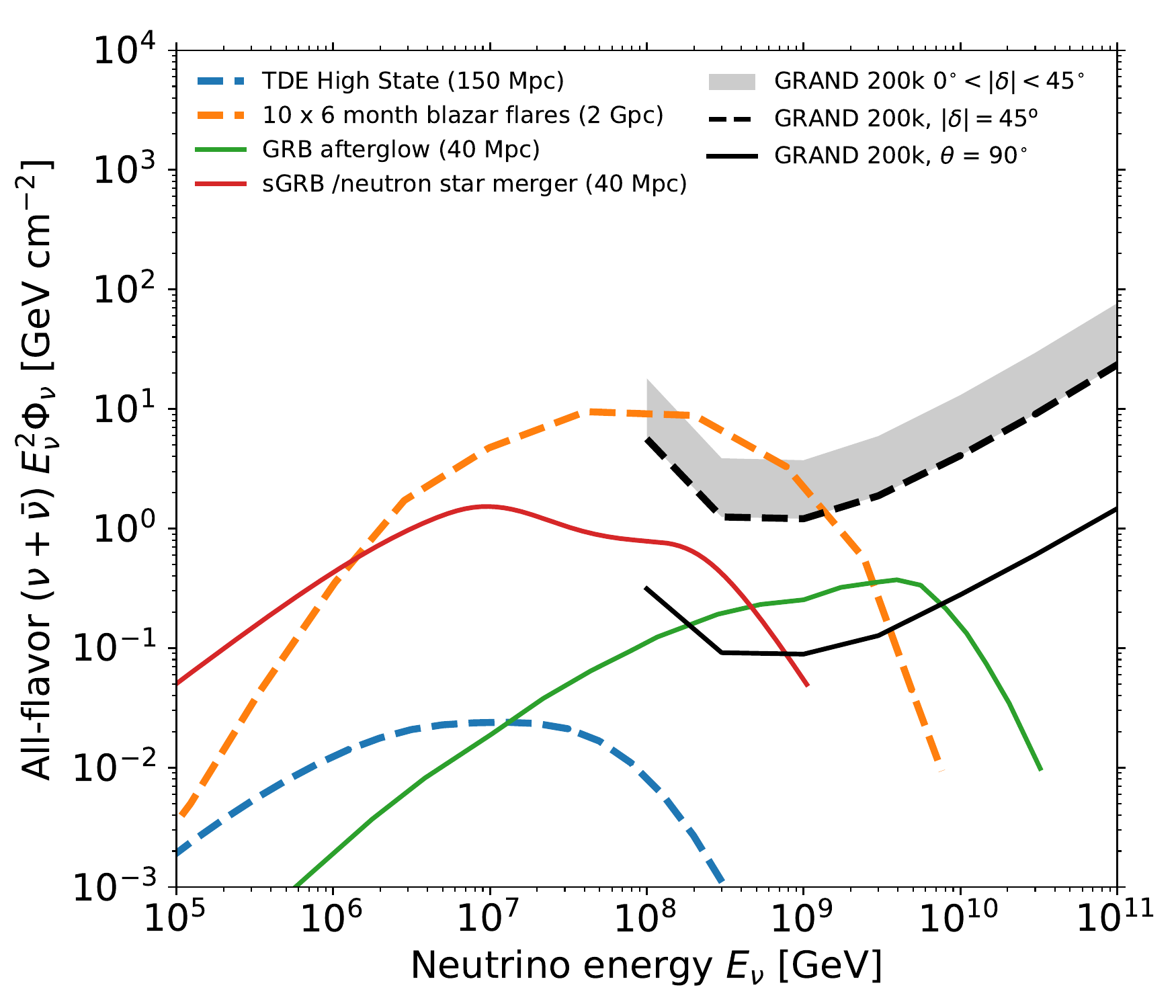}
\caption{{\it Left:} Differential and integrated neutrino sensitivity limits calculated from the $10,000$ antennas simulation ("GRAND10k", pink area) and the extrapolation for the 20-times larger GRAND array ("GRAND200k", maroon line). The gray region represents the all-flavor cosmogenic neutrinos flux expectations derived from the results of the Pierre Auger Observatory \cite{AlvesBatista:2018zui}. Adapted from \cite{GRAND20}. {\it Right:} GRAND point source sensitivity limits \cite{GRAND20}. Short-duration transients (short GRBs, GRB afterglows) are compared to the GRAND200k instantaneous sensitivity at zenith angle $\theta = 90^\circ$ (solid black line). Long-duration transients (e.g., TDE) are compared to declination-averaged sensitivity (gray-shaded band). The stacked fluence from 10 six-month-long blazar flares in the declination range $40^\circ< |\delta| < 45^\circ$ is compared to the GRAND200k sensitivity for a fixed $\delta=45^\circ$ (dashed black line). The GRAND limits assume that the 200k antennas are deployed at a single location.  
}
\label{fig:nu_performances}
\end{figure}

The development of a GRAND end-to-end simulation chain and of several reconstruction tools dedicated to inclined EAS have enabled to assess the performances of GRAND for UHE neutrino, cosmic ray, and gamma-ray detection. The simulation chain comprises a 3-D Monte-Carlo sampler of tau leptons generated by $\nu_\tau$ interactions underground (DANTON \cite{DANTON:note}), a semi-analytical radio-signal fast computation tool (Radio-Morphing \cite{Zilles20_RM,Chiche_ICRC21}), and an antenna response module (NEC4 \cite{NEC4}). The final step is the detector trigger simulation. Our trigger condition requires $\ge 5$ units in one 9-antenna square cell to be triggered, and the peak-to-peak amplitude of the voltage signal at the output of the antennas to be $\ge 30 [75] \mu$V (twice the expected stationary background noise in the $50-200$\,MHz frequency range) in the aggressive [conservative] scenario. 

This simulation chain was run over a $10,000$\,km$^2$ area, with $10,000$ antennas deployed along a square grid of 1 km step size in a basin surrounded by high peaks of the TianShan mountain range in China. The 10-year 90\% C.L. GRAND sensitivity limit (Fig.~\ref{fig:nu_performances}, {\it left}) is scaled from the simulated region to $200,000$ km$^2$ (GRAND200k). The integrated limits correspond to the Feldman-Cousins upper limit per decade in energy at 90\% C.L., assuming a power-law neutrino spectrum $\propto E_{\nu}^{-2}$, for no candidate events and null background. The 10-year GRAND integrated sensitivity limit is $\sim 10^{-10}\,{\rm GeV}\,{\rm cm}^{-2}\,{\rm s}^{-1}\,{\rm sr}^{-1}$ above $5\times 10^{17}$\,eV \cite{GRAND20}. 

\begin{figure}[t]
\centering
\includegraphics[width=0.48\linewidth]{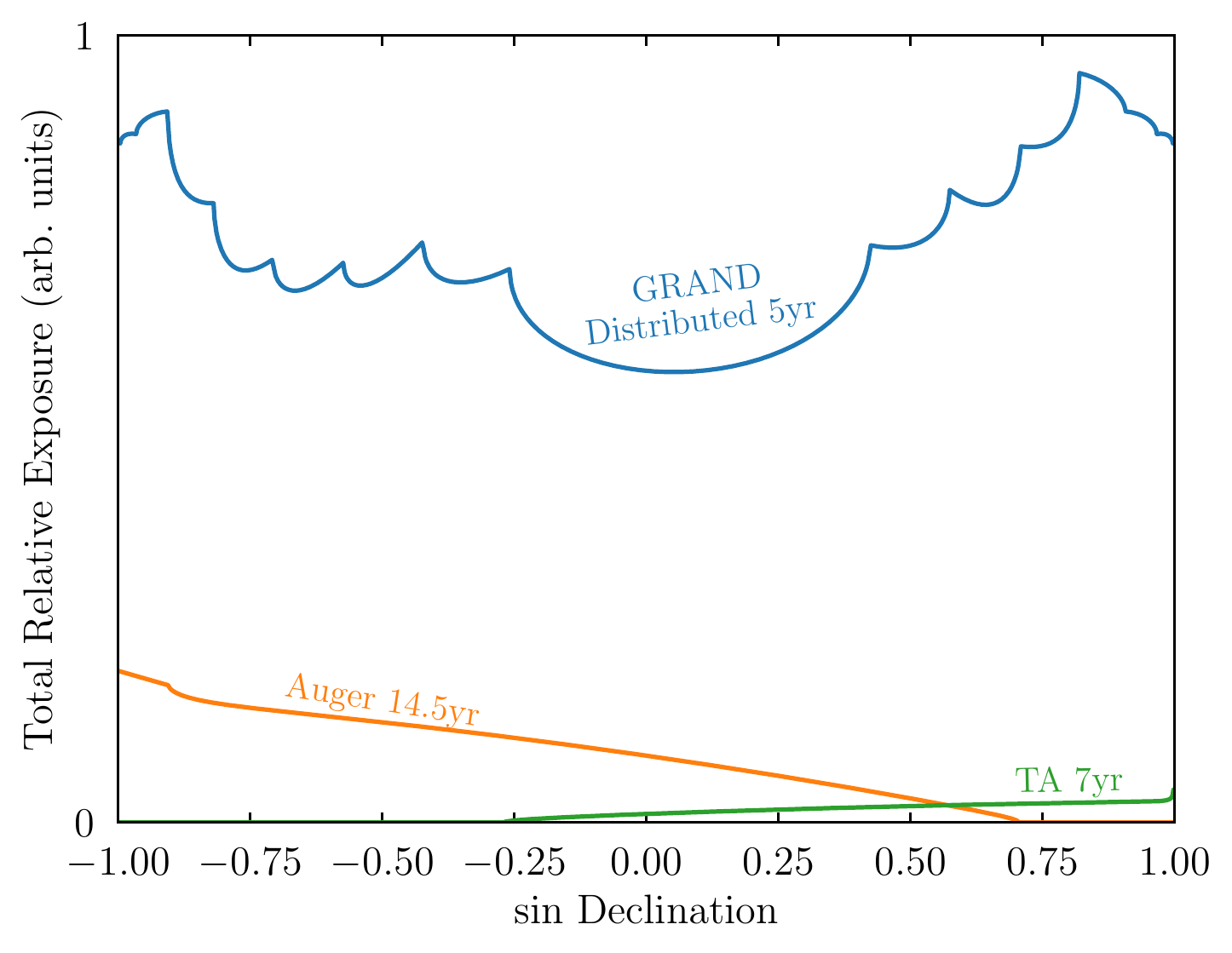}
\includegraphics[width=0.48\linewidth]{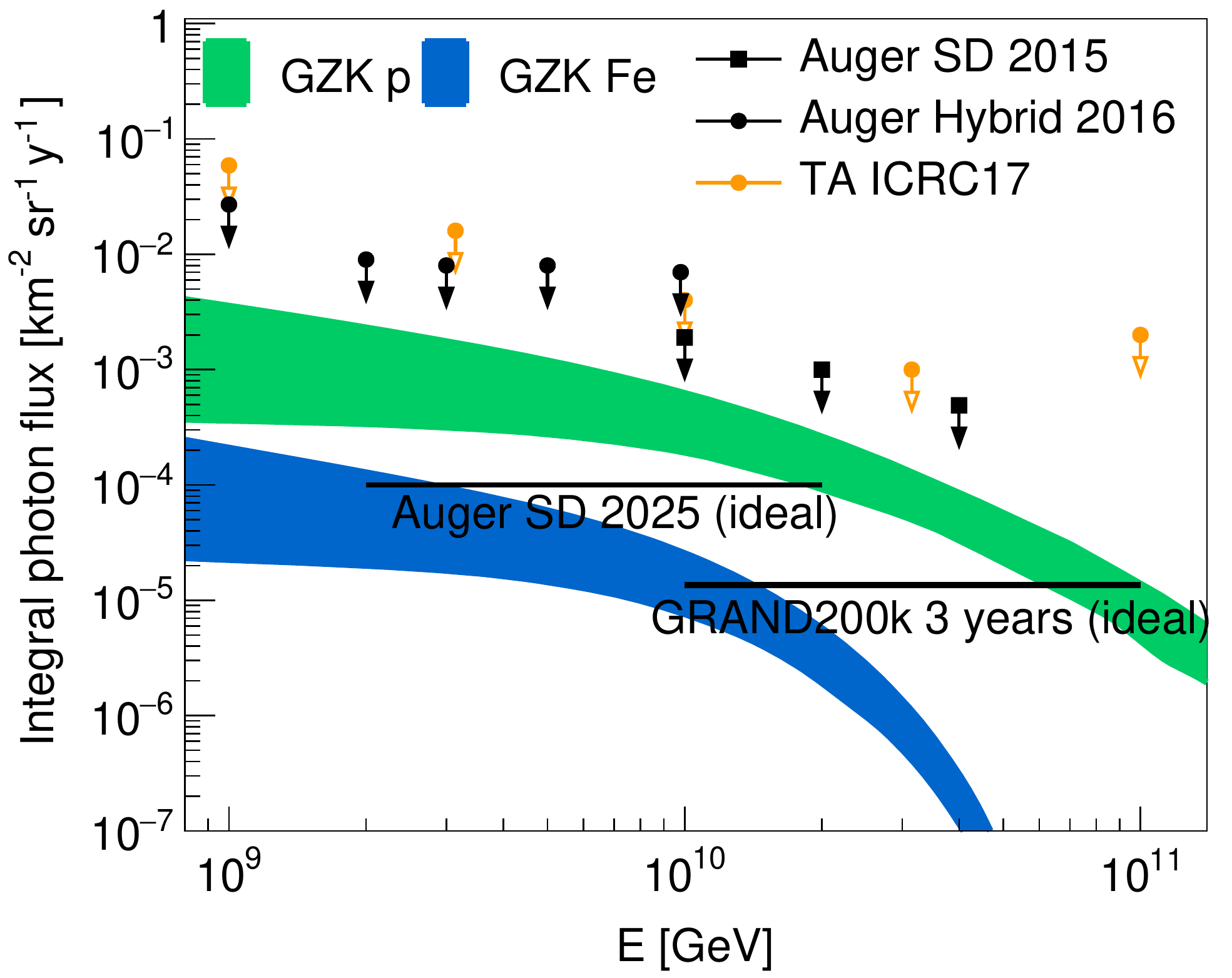}
\caption{{\it Left:} The relative annual geometric exposure to UHECRs of GRAND for a uniform distribution of 10 sub-arrays on the Earth, compared to Auger and the Telescope Array (TA). {\it Right:} Projected upper limits of GRAND on UHE photon sensitivity after 3 years of operation. For comparison, we include the existing upper limits from Auger and TA, and the projected reach of Auger by 2025. Overlayed are the predicted cosmogenic UHE photon flux from pure-proton and pure-iron UHECRs, as estimated in \cite{Sarkar11}. }
\label{fig:CR_gamma_performances}
\end{figure}

For UHECR detection, GRAND will be fully efficient above $10^{18}$\,eV and sensitive to cosmic rays in a zenith-angle range of $65^\circ-85^\circ$. The geometrical aperture of the experiment will be $107,000\,$km$^2$\,sr. However, when including events with shower cores outside the instrumented area and when taking trigger conditions into account, UHECR air-shower simulations indicate that GRAND would have a $4-5$ times higher exposure. Figure~\ref{fig:CR_gamma_performances} ({\it left}) presents an example of the GRAND exposure to UHECR detection, assuming 10 random locations of sub-arrays of $20,000$ antennas uniformly spaced between geographical latitudes 60N and 40S. An uniform acceptance was assumed over zenith angles of $65^\circ-85^\circ$. A full-sky coverage is obtained with such a configuration.

The aperture of GRAND to UHE gamma rays is similar to the one of UHECRs. Figure~\ref{fig:CR_gamma_performances} ({\it right}) shows that the sensitivity of GRAND200k to UHE gamma rays is sufficient to detect them even in the pessimistic case where UHECRs are heavy. To compute the preliminary sensitivity of GRAND200k to UHE gamma rays, we assumed that the detector is fully efficient to gamma ray-initiated air showers with energies above $10^{10}$~GeV in the zenith range $60^{\circ}$--$85^{\circ}$.  The sensitivity shown is the Feldman-Cousins upper limit at the $95\%$ C.L., assuming no candidate events, null background, and a UHE gamma-ray spectrum $\propto E^{-2}$.  The assumption of a background-free search is reasonable in the $10^{10}$--$10^{10.5}$~GeV range, even for the conservative hypothesis that GRAND reaches a resolution in $X_{\rm max}$ of only 40~g~cm$^{-2}$. 

Novel reconstruction methods performing fits to the strength of the radio signal as a function of the angle from the shower axis (angular distribution function) have demonstrated that angular resolutions of $\sim 0.1^\circ$ could be achieved on the particle arrival direction \cite{DecoenePhD, Decoene_ICRC2021}, rendering neutrino and gamma-ray astronomy possible with GRAND. 
For a given sub-array location, the instantaneous neutrino field of view of GRAND is a band between zenith angles $85^\circ{\leq}\theta {\leq}95^\circ$, 
corresponding to ${<}5\%$ of the sky.  
Assuming that all azimuth angles are observed at any instant, 
approximately 80\% of the sky is observed every day by each sub-array. With $10-20$ locations spread around the globe, GRAND will offer a continuous full-sky coverage which enables multi-messenger astronomy in combination with any other experiment on Earth or in space.

Preliminary results obtained on the energy resolution are encouraging, as expected generally for energy reconstruction with radio measurements. A preliminary reconstruction method using the radio signal lateral distribution function, with no detector response implemented, leads to a 4\% energy resolution. Another preliminary global reconstruction method using the angular distribution function leads to a 20\% energy resolution \cite{Decoene_ICRC2021}. Hence a final energy resolution of 10\% is likely to be achieved. 
Finally, resolutions on $X_{\rm max}$ better than $40\,{\rm g}\,{\rm cm}^{-2}$ were achieved in preliminary studies based on \cite{Buitink:2014eqa}. More refined and optimized methods are being developed to improve the reconstruction of all EAS parameters.

\section{A rich science case}

GRAND ambitions to tackle a variety of long-standing astrophysics and fundamental physics questions. We list the major questions on which GRAND has a potential to make breakthroughs. 

\paragraph{Diffuse neutrino fluxes.} With an increase of almost two decades in neutrino sensitivity compared to existing experiments, GRAND ensures the detection of EeV neutrinos. Cosmogenic neutrino studies show that the results from GRAND should severely constrain the sources of UHECRs whatever the outcome of the measurements \cite{AlvesBatista:2018zui,Moller2019}, and constrain the proton fraction at UHE \cite{van_Vliet_2019}. The GRAND sensitivity, combined with its sub-degree angular resolution, will open the possibility to perform UHE neutrino astronomy, by identifying point-sources \cite{Fang:2016hop}. Note that the sources of UHECRs and UHE neutrinos could be different: transparent source environments are indeed favored to let UHECR escape from the sources, while thicker environments could lead to more abundant neutrino production. Hence, even if a heavy composition was measured for observed UHECRs, it would not necessarily imply that the flux of neutrinos at EeV should be suppressed.

\paragraph{Transient EeV neutrino astronomy.} Thanks to its sub-degree angular resolution and its full-sky coverage, GRAND could identify EeV neutrino sources by detecting neutrinos from transient events in coincidence with electromagnetic emission \cite{Guepin:2017dfi, 2020PhRvD.102l3013V}. Figure~\ref{fig:nu_performances} ({\it right}) compares theoretical neutrino fluence estimates from transient sources to the GRAND point-source sensitivities. We present a short-duration gamma-ray burst (sGRB) possibly associated with a double neutron-star merger \cite{Kimura_2017} at 40~Mpc and a GRB afterglow\ \cite{Murase:2007yt} at 40~Mpc, a tidal disruption event (TDE) at 150~Mpc\ \cite{Guepin:2017abw}, and the stacked fluence of 10 blazar flares in the declination range $40^\circ< |\delta| < 45^\circ$, calculated using as template a 6-month long flare of the blazar 3C66A at 2~Gpc\ \cite{Murase:2014foa}.
The sources were assumed to lie at distances such to allow for a conservative rate of $\sim$1 event per century.
Depending on the background discrimination efficiency, GRAND will be able send alerts to other experiments or coordinated systems like AMON~\cite{SOLARES2019} for follow-up campaigns.

\paragraph{UHECR and gamma rays.} According to preliminary simulations, GRAND will have full detection efficiency for cosmic rays with zenith angles larger than $70^\circ$ and energies above $10^{18}$\,eV \cite{GRAND20}. This will yield an exposure $\gtrsim 15$ times larger than the Pierre Auger Observatory. Further, it would be a full-sky instrument, which is crucial to study anisotropy \cite{Denton_2015}. 

Assuming that an $X_{\rm max}$ resolution of $40\,{\rm g}\,{\rm cm}^{-2}$ is achieved --a realistic goal given present experimental results \cite{Buitink:2016nkf,Bezyazeekov:2015rpa} and preliminary simulations results (see Section~\ref{section:performances})--, GRAND will be able to distinguish between UHECR and UHE gamma-ray showers. The non-detection of cosmogenic gamma rays within 3 years of operation of GRAND would exclude a light composition of UHECRs, while a detection of UHE gamma rays from nearby sources would probe the cosmic radio background \cite{Fixsen:1998kq}.

\paragraph{Fundamental physics.} High-energy cosmic neutrinos provide a chance to test fundamental physics in new regimes~\cite{2019BAAS...51c.215V}.
Numerous new-physics models have effects whose intensities are proportional to some power of the neutrino energy and to the source-detector baseline. GRAND could probe new physics with exquisite sensitivities, see e.g., \cite{2020PhRvD.102l3019D}, and will be able to test dark matter models through neutrino and photon constraints.

\paragraph{Transient radio-astronomy.} By incoherently adding the signals from the large number antennas in a subarray, GRAND will also be able to detect a 30-Jy fast radio burst (FRB) with a flat frequency spectrum \cite{GRAND20}. As incoherent summing preserves the wide field of view of a single antenna, GRAND may be able to detect several hundreds of FRBs per day. In addition, the detection of a single FRB by several sub-arrays would enable to reconstruct the arrival direction of the radio signal.

\section{Technical challenges}

Autonomous radio-detection, i.e., identifying EAS radio signals with radio antennas alone, is a major challenge due to the ubiquitously dominant radio background, which necessitates an important rejection efficiency. It has been shown that EAS radio signatures differ from background events, with much shorter time traces \cite{Barwick:2016mxm} and specific amplitude \cite{Nelles:2014dja} and polarization patterns at ground \cite{carduner2017codalemaextasis}. These unique features have been used by the TREND-experiment to perform an efficient rejection of the background signals using radio data only \cite{Charrier:2018fle}. An efficient background rejection criterion ($\sim 99.9\%$ of noise-induced event rejection) at the data acquisition level has also been developed recently, based on the projection of the total electric field along the direction of the local magnetic field \cite{Chiche_ICRC21}. Interestingly, this criterion can also be used to discriminate neutrino and cosmic-ray EAS. These methods are being refined, and sophisticated data treatment techniques (adaptative filtering, machine learning, etc.) are being developed in parallel \cite{2018arXiv180901934F,2019JInst..14P4005E}.

From a hardware point-of-view, an adequate DAQ system, which can treat signals at a high-enough frequency rate ($\sim$\,kHz) should enable to perform efficient triggering. This system has been implemented in the first 300-antenna prototype, GRANDProto300 (GP300) \cite{GP300_ICRC2021}. The prototype will serve as a test bench to validate solutions for the next stages of GRAND. It will be deployed over 200\,km$^2$ in an environment with excellent radio quality, presenting a low rate of transient radio pulses in particular. The protocol used in the site survey for the GP300 phase of GRAND will be extended and optimized when validating the locations of the GRAND sub-arrays.

\section{The road to neutrino astronomy}

GRAND will be modular and built in stages. Between 2021 and 2025, the 300-antenna pathfinder, GP300, will validate the GRAND detection principle, test and optimize the detection units design, the autonomous trigger and data transfer strategies. GP300 will also conduct an ambitious science program on cosmic rays between $10^{16.5-18}\,$eV \cite{GP300_ICRC2021}. $10,000$ detection units of the finalized design will be produced and deployed in 2025 to create GRAND10k, the first GRAND sub-array. This array will serve to test challenges related to large-scale arrays, such as communication and data transfer/storage. GRAND10k likely has the sensitivity to detect the first EeV neutrinos. By the 2030s, once this first sub-array has been demonstrated to operate successfully, its design will be frozen. Industrial companies will be prospected to replicate this sub-array and take care of the mass-production and deployment of the units with predefined specifications in terms of reliability, costs etc. The design of each sub-array may be adapted, depending on location and topography, or to address specific science cases.

\bibliographystyle{elsarticle-num}
{\footnotesize
\bibliography{ICRC}
}
\clearpage
\section*{Full Authors List: \Coll\ Collaboration}

%
%

Jaime \'Alvarez-Mu\~niz$^{\ref{Santiago}}$,
Rafael Alves Batista$^{\ref{Radboud}}$,
Aur\'elien Benoit-L\'evy$^{\ref{LIST}}$,
Julien Bolmont$^{\ref{LPNHE}}$,
Henk Brans$^{\ref{Radboud}}$,
Mauricio Bustamante$^{\ref{NBI}}$, 
Didier Charrier$^{\ref{SUBATECH}}$,
LingMei Cheng$^{\ref{NAOC}}$,
Simon Chiche$^{\ref{IAP}}$,
Zigao Dai$^{\ref{Anhui}}$,
Rogerio M. de Almeida$^{\ref{UFF}}$,
Valentin Decoene$^{\ref{PSU}}$,
Peter B.\ Denton$^{\ref{BNL}}$, 
Beatriz de Errico$^{\ref{UFRJ}}$,
Sijbrand De Jong$^{\ref{Radboud},\ref{Nikhef}}$,
João R. T. de Mello Neto$^{\ref{UFRJ}}$,
Krijn D.\ De Vries$^{\ref{IIHE}}$,
Kaikai Duan$^{\ref{PMO}}$,
Ran Duan$^{\ref{NAOC}}$,
Ralph Engel$^{\ref{KIT_IKP},\ref{KIT_ETP}}$,
Yizhong Fan$^{\ref{PMO}}$,
Ke Fang$^{\ref{WIPAC}}$,
QuanBu Gou$^{\ref{IHEP}}$,
Junhua Gu$^{\ref{NAOC}}$,
Claire Gu\'epin$^{\ref{Maryland},\ref{JSSI}}$,
Jianhua Guo$^{\ref{PMO}}$,
Yiqing Guo$^{\ref{IHEP}}$,
Rene Habraken$^{\ref{Radboud},\ref{Nikhef}}$,
Andreas Haungs$^{\ref{KIT_IKP}}$,
Haoning He$^{\ref{PMO}}$,
Eric Hivon$^{\ref{IAP}}$,
Hongbo Hu$^{\ref{IHEP}}$,
Xiaoyuan Huang$^{\ref{PMO}}$,
Yan Huang$^{\ref{NAOC}}$,
Tim Huege$^{\ref{KIT_IKP},\ref{VUB_AI}}$,
Marcelo Ismerio Oliveira$^{\ref{UFRJ}}$,
Ramesh Koirala$^{\ref{NanjingU1},\ref{NanjingU2}}$,
Kumiko Kotera$^{\ref{IAP},\ref{VUB_ELEM}}$,
Wen Jiang$^{\ref{Xidian}}$,
Bruno L. Lago$^{\ref{Fonseca}}$,
Sandra Le Coz$^{\ref{LPNHE}}$, 
Jean-Philippe Lenain$^{\ref{LPNHE}}$,
Bo Liu$^{\ref{Xidian}}$,
Cheng Liu$^{\ref{IHEP}}$,
Ruoyu Liu$^{\ref{NanjingU1},\ref{NanjingU2}}$,
Wei Liu$^{\ref{IHEP}}$,
Pengxiong Ma$^{\ref{PMO}}$,
Olivier Martineau-Huynh$^{\ref{LPNHE},\ref{NAOC}, \ref{IAP}}$,
Miguel Mostaf\'a$^{\ref{PSU_part},\ref{PSU}}$,
Fabrice Mottez$^{\ref{LUTH}}$, 
Jean Mouette$^{\ref{IAP}}$, 
Kohta Murase$^{\ref{PSU_part},\ref{PSU}}$,
Valentin Niess$^{\ref{Clermont}}$,
Foteini Oikonomou$^{\ref{UTrondheim}}$,
Ziwei Ou$^{\ref{SunYatsen}}$,
Tanguy Pierog$^{\ref{KIT_IKP}}$,
Lech Wiktor Piotrowski$^{\ref{UWarsaw}}$,
Simon Prunet$^{\ref{Lagrange}}$,
Xiangli Qian$^{\ref{SDMU}}$,
Inge van Rens$^{\ref{Radboud}}$,
Valentina Richard Romei$^{\ref{IAP}}$,
Markus Roth$^{\ref{KIT_IKP}}$,
Fabian Sch\"ussler$^{\ref{IRFU}}$, 
Dániel Szálas-Motesiczky$^{\ref{Radboud}}$, 
Jikke Tacken$^{\ref{Radboud}}$, 
Anne Timmermans$^{\ref{Radboud},\ref{IIHE}}$,
Charles Timmermans$^{\ref{Radboud},\ref{Nikhef}}$,
Mat\'ias Tueros$^{\ref{CONICET},\ref{IAP}}$,
Rongjuan Wang$^{\ref{Xidian}}$,
Shen Wang$^{\ref{PMO}}$,
Xiangyu Wang$^{\ref{NanjingU1},\ref{NanjingU2}}$,
Xu Wang$^{\ref{Shandong}}$,
Clara Watanabe$^{\ref{UFRJ}}$,
Daming Wei$^{\ref{PMO}}$,
Feng Wei$^{\ref{Xidian}}$,
Thei Wijnen$^{\ref{Radboud}}$,
Xiangping Wu$^{\ref{NAOC},\ref{Shanghai}}$,
Xuefeng Wu$^{\ref{PMO2}}$,
Xin Xu$^{\ref{Xidian}}$,
Xing Xu$^{\ref{PMO}}$,
Lili Yang$^{\ref{SunYatsen}}$,
Xuan Yang$^{\ref{PMO}}$,
Qiang Yuan$^{\ref{PMO}}$,
Philippe Zarka$^{\ref{LESIA}}$,
Houdun Zeng$^{\ref{PMO}}$,
Bing\ Theodore Zhang$^{\ref{PSU}}$,
Chao Zhang$^{\ref{KIT_IKP}, \ref{KIAA}, \ref{PekingU}}$,
Jianli Zhang$^{\ref{NAOC}}$,
Kewen Zhang$^{\ref{LPNHE}}$,
Pengfei Zhang$^{\ref{Xidian}}$,
Songbo Zhang$^{\ref{PMO2}}$,
Yi Zhang$^{\ref{PMO},\ref{IHEP}}$,
Hao Zhou$^{\ref{JiaoTong}}$\\

\small

\begin{affil}{}\label{Santiago}
Departamento de F\'isica de Part\'iculas \& Instituto Galego de F\'\i sica de Altas Enerx\'\i as, Universidad de Santiago de Compostela, 15782 Santiago de Compostela, Spain 
\end{affil}

\begin{affil}{}\label{Radboud}
Institute for Mathematics, Astrophysics and Particle Physics (IMAPP), Radboud Universiteit, Nijmegen, Netherlands
\end{affil}

\begin{affil}{}\label{LIST}
Université Paris-Saclay, CEA, List, F-91120, Palaiseau, France
\end{affil}

\begin{affil}{}\label{LPNHE}
Sorbonne Universit\'e, Universit\'e Paris Diderot, Sorbonne Paris Cit\'e, CNRS, Laboratoire de Physique Nucl\'eaire et de Hautes Energies (LPNHE), 4 place Jussieu, F-75252, Paris Cedex 5, France 
\end{affil}

\begin{affil}{}\label{NBI}
Niels Bohr International Academy, Niels Bohr Institute, 2100 Copenhagen, Denmark
\end{affil}

\begin{affil}{}\label{SUBATECH}
SUBATECH, Institut Mines-Telecom Atlantique -- CNRS/IN2P3 -- Universit\'e de Nantes, Nantes, France
\end{affil}

\begin{affil}{}\label{NAOC}
National Astronomical Observatories, Chinese Academy of Sciences, Beijing 100101, China
\end{affil}

\begin{affil}{}\label{IAP}
Sorbonne Universit\'e, CNRS, UMR 7095, Institut d'Astrophysique de Paris, 98 bis bd Arago, 75014 Paris, France
\end{affil}

\begin{affil}{}\label{Anhui}
University of Science and Technology of China, 230026 Hefei, Anhui, China
\end{affil}

\begin{affil}{}\label{VUB_AI}
Astrophysical Institute, Vrije Universiteit Brussel, Pleinlaan 2, 1050 Brussel, Belgium
\end{affil}

\begin{affil}{}\label{UFF}
Universidade Federal Fluminense, EEIMVR, Volta Redonda, RJ, Brazil
\end{affil}

\begin{affil}{}\label{PSU}
Department of Physics, Department of Astronomy \& Astrophysics, Pennsylvania State University, University Park, PA 16802, USA
\end{affil}

\begin{affil}{}\label{BNL}
High Energy Theory Group, Physics Department,Brookhaven National Laboratory, Upton, NY 11973, USA
\end{affil}

\begin{affil}{}\label{UFRJ}
Universidade  Federal  do  Rio  de  Janeiro  (UFRJ),  Instituto  de  F\'isica,  Brazil
\end{affil}

\begin{affil}{}\label{Nikhef}
Nationaal Instituut voor Kernfysica en Hoge Energie Fysica (Nikhef), Netherlands
\end{affil}

\begin{affil}{}\label{IIHE}
IIHE/ELEM, Vrije Universiteit Brussel, Pleinlaan 2, 1050 Brussels, Belgium
\end{affil}

\begin{affil}{}\label{KIT_IKP}
Institute for Astroparticle Physics, Karlsruhe Institute of Technology (KIT), D-76021 Karlsruhe, Germany
\end{affil}

\begin{affil}{}\label{KIT_ETP}
Institute of Experimental Particle Physics (ETP), Karlsruhe Institute of Technology (KIT), D-76021 Karlsruhe, Germany
\end{affil}

\begin{affil}{}\label{Maryland}
Department of Astronomy, University of Maryland, College Park, MD 20742-2421, USA
\end{affil}

\begin{affil}{}\label{JSSI}
Joint Space-Science Institute, College Park, MD 20742-2421, USA
\end{affil}

\begin{affil}{}\label{PMO}
Key Laboratory of Dark Matter and Space Astronomy, Purple Mountain Observatory, Chinese Academy of Sciences,210023 Nanjing, Jiangsu, China
\end{affil}

\begin{affil}{}\label{WIPAC}
Wisconsin IceCube Particle Astrophysics Center (WIPAC) and Dept. of Physics, University of Wisconsin-Madison, Madison, WI 53703, USA
\end{affil}

\begin{affil}{}\label{IHEP}
Institute of High Energy Physics, Chinese Academy of Sciences, 19B YuquanLu, Beijing 100049, China
\end{affil}

\begin{affil}{}\label{Xidian}
Key Laboratory of Antennas and Microwave Technology, Xidian University, Xi’an 710071, China
\end{affil}

\begin{affil}{}\label{NanjingU1}
School of Astronomy and Space Science, Xianlin Road 163, Nanjing University, Nanjing 210023, China
\end{affil}

\begin{affil}{}\label{NanjingU2}
Key laboratory of Modern Astronomy and Astrophysics (Nanjing University), Ministry of Education, Nanjing 210023, People’s Republic of China
\end{affil}

\begin{affil}{}\label{VUB_ELEM}
Vrije Universiteit Brussel (VUB), Dienst ELEM, Pleinlaan 2, B-1050, Brussels, Belgium
\end{affil}

\begin{affil}{}\label{Fonseca}
Centro Federal de Educação Tecnológica Celso Suckow da Fonseca, Nova Friburgo, Brazil
\end{affil}

\begin{affil}{}\label{PSU_part}
Center for Multimessenger Astrophysics, Pennsylvania State University, University Park, PA 16802, USA
\end{affil}

\begin{affil}{}\label{LUTH}
LUTH, Obs.\ de Paris, CNRS, Universit\'e Paris Diderot, PSL Research University, 5 place Jules Janssen, 92190 Meudon, France
\end{affil}

\begin{affil}{}\label{Clermont}
Université Clermont Auvergne, CNRS/IN2P3, LPC, F-63000 Clermont-Ferrand, France.
\end{affil}
 
\begin{affil}{}\label{UTrondheim}
Institutt for fysikk, NTNU, Trondheim, Norway
\end{affil}

\begin{affil}{}\label{SunYatsen}
School of Physics and Astronomy, Sun Yat-sen University, Zhuhai 519082, China
\end{affil}

\begin{affil}{}\label{UWarsaw}
Faculty of Physics, University of Warsaw, Pasteura 5, 02-093 Warsaw, Poland
\end{affil}

\begin{affil}{}\label{Lagrange}
Laboratoire Lagrange, Universit\'e Côte d'Azur, Observatoire de la Côte d’Azur, CNRS, Parc Valrose, 06104 Nice Cedex 2, France
\end{affil}

\begin{affil}{}\label{SDMU}
Department of Mechanical and Electrical Engineering, Shandong Management University, Jinan 250357, China 
\end{affil}

\begin{affil}{}\label{IRFU}
IRFU, CEA, Universit\'e Paris-Saclay, F-91191 Gif-sur-Yvette, France
\end{affil}

\begin{affil}{}\label{CONICET}
Instituto de F\'isica La Plata, CONICET, Boulevard 120 y 63 (1900), La Plata, Argentina
\end{affil}

\begin{affil}{}\label{Shandong}
Department of Mechanical and Electrical Engineering, Shandong Management University, Jinan 250357,China.
\end{affil}

\begin{affil}{}\label{Shanghai}
Shanghai Astronomical Observatory, Chinese Academy of Sciences, 80 Nandan Road, Shanghai 200030, China 
\end{affil}

\begin{affil}{}\label{PMO2}
Purple Mountain Observatory, Chinese Academy of Sciences, Nanjing 210023, China
\end{affil}

\begin{affil}{}\label{LESIA}
LESIA, Observatoire de Paris, CNRS, PSL/SU/UPD/SPC, Place J. Janssen, 92195 Meudon, France
\end{affil}

\begin{affil}{}\label{KIAA}
Kavli Institute for Astronomy and Astrophysics, Peking University, Beijing 100871, China
\end{affil}

\begin{affil}{}\label{PekingU}
Department of Astronomy, School of Physics, Peking University, Beijing 100871, China 
\end{affil}

\begin{affil}{}\label{JiaoTong}
Tsung-Dao Lee Institute \& School of Physics and Astronomy, Shanghai Jiao Tong University, 200240 Shanghai,China
\end{affil}

\end{document}